\documentclass[fleqn,usenatbib]{mnras}

\usepackage{newtxtext,newtxmath}

\usepackage[T1]{fontenc}
\usepackage{ae,aecompl}

\usepackage{graphicx}	
\usepackage{amsmath}	
\usepackage{amssymb}	

\newcommand{\source}{PKS~B1322$-$110\,}
\newcommand{\upd}[1]{{#1}}
\newcommand{\bea}{\begin{eqnarray}}
\newcommand{\eea}{\end{eqnarray}}

\title[Annual cycle of PKS B1322]{Spica and the annual cycle of PKS B1322$-$110 scintillations}

\author[H. Bignall et al.]{Hayley Bignall,$^{1}$\thanks{E-mail: Hayley.Bignall@csiro.au}
Cormac Reynolds,$^{1}$
Jamie Stevens,$^{2}$
Keith Bannister,$^{3}$
\newauthor
Simon Johnston,$^{3}$
Artem V. Tuntsov,$^{4}$\thanks{E-mail: Artem.Tuntsov@manlyastrophysics.org}
Mark A. Walker,$^{4}$
Sergei Gulyaev,$^{5}$
\newauthor
Tim Natusch,$^{5}$
Stuart Weston,$^{5}$
Noor Masdiana Md Said,$^{6}$
Matthew Kratzer.$^{7}$\\
\\
$^{1}$CSIRO Astronomy and Space Science, 26 Dick Perry Avenue, Kensington, WA 6151, Australia\\
$^{2}$CSIRO Paul Wild Observatory, 1828 Yarrie Lake Road, Narrabri, NSW 2390, Australia\\
$^{3}$CSIRO Astronomy and Space Science, PO Box 76, Epping NSW 1710, Australia\\
$^{4}$Manly Astrophysics, 15/41-42 East Esplanade, Manly, NSW 2095, Australia\\
$^{5}$Auckland University of Technology, 120 Mayoral Drive, Auckland 1010, New Zealand\\
$^{6}$University of Tasmania, Hobart, TAS 7001, Australia\\
$^{7}$The University of Queensland, QLD 4072, Australia
}

\date{Accepted 2019 May 31. Received 2019 May 06; in original form 2018 December 07}

\pubyear{2019}

\begin{document}
\label{firstpage}
\pagerange{\pageref{firstpage}--\pageref{lastpage}}
\maketitle

\begin{abstract}
PKS B1322$-$110 is a radio quasar that is located only $8^\prime\!.5$ in angular separation from the bright B star Spica. It exhibits intra-day variability in its flux density at GHz frequencies attributed to scintillations from plasma inhomogeneities. We have tracked the rate of scintillation of this source for over a year with the Australia Telescope Compact Array, recording a strong annual cycle that includes a near-standstill in August and another in December. The cycle is consistent with scattering by highly anisotropic plasma microstructure, and we fit our data to that model in order to determine the kinematic parameters of the plasma. Because of the low ecliptic latitude of \source, the orientation of the plasma microstructure is poorly constrained. Nonetheless at each possible orientation our data single out a narrow range of the corresponding velocity component, leading to a one-dimensional constraint in a two-dimensional parameter space. The constrained region is consistent with a published model in which the scattering material is associated with Spica and consists of filaments that are radially oriented around the star. This result has a 1\%\ probability of arising by chance.
\end{abstract}




\section{Introduction}
\label{section:intro}

\source is a flat spectrum radio source \citep{cataloguepaper} which was recently discovered to undergo extreme intra-day variations (IDV, also used to refer to the class of sources) at GHz frequencies \citep{telegram}. Rapid flux density variations in these sources are scintillations resulting from plasma inhomogeneities in our Galaxy \citep{jaunceyetalreview, rickettreview}, and can be used to constrain the apparent size or brightness temperature of their radio emitting regions \citep{muasquasar,discovery0405}.

Scintillation can be thought of as a spatial flux pattern -- i.e. the source projected through the transparent plasma ``screen'' -- that drifts relative to the Earth. For sources at cosmological distances, the velocity of the pattern is essentially that of the screen \citep{effectivevelocity}, and therefore the change in the velocity of the Earth as it orbits the Sun can strongly affect the variation timescales. This annual modulation has so far been reported in a handful of sources: J1819+3845 \citep{cycle1819,cycle1819aa}; QSO B0917+624 \citep{jaunceymacquart, rickettetal,cycle0917but}; PKS~1257$-$326 \citep{cycle1257}; PKS~B1519$-$273 \citep{cycle1519};  PKS~B1622$-$253 \citep{cycle15191622}; S5~0716+714 \citep{cycle0716}; 0925+504 \citep{cycle0925apss, cycle0925mnras}; S4~0954+65 \citep{cycle0954}; 1156+295 (4C+29.45,  \citealt{cycle1156}); and J1128+5925 \citep{cycle1128, cycle1128but}.  Together with the two-station experiments \citep{twostation1819, twostation1257}, which directly demonstrated the existence of a spatial flux pattern, the annual cycles provided the key evidence that proved IDV to be scintillation. No clear annual cycle could be established for the prototypical intra-hour variable PKS~0405$-$385 due to the intermittency of its variations \citep{nocycle0405}.

Annual cycles of the two best-studied intra-hour variables, J1819+3845 and PKS~1257$-$326 have been shown to be consistent with highly anisotropic, essentially one-dimensional scattering \citep{walkerdebruynbignall}, and the orientation of the respective anisotropy axes along with the projected velocity of the screen were determined. However, although the properties of the screens have been precisely characterised, the physical context of the scattering material remains unclear. In the case of J1819+3845, whose screen is expected to be relatively close, less than $10\,\mathrm{pc}$ from Earth, it was previously suggested that the plasma might be associated with Vega, a nearby A star that is close to the source in the sky \citep{twostation1819}. Curiously, the anisotropy axis of J1819+3845 does point towards Vega.

The possibility of a connection between IDVs and hot stars was reinforced by the realisation that the new IDV \source\ is just 8.5 arcminutes away from Spica, the Sun's closest B star neighbour, prompting \citet{walkeretal2017} to examine the stars foreground to J1819+3845 and PKS~B1257$-$326. A conclusion of that study was that the scintillations of both sources are due to plasma associated with nearby A stars --- the star being Alhakim ($\iota$ Cen), in the case of PKS~B1257$-$326. The picture that was suggested by \citet{walkeretal2017} is of plasma filaments that are radially oriented around the host star, and co-moving with it.

The possible connection between Spica and the IDV of \source\ was left out of that analysis. The reason for the omission is that the close alignment between Spica and \source\ motivated the idea of an association, and therefore cannot be used as a test. On the other hand, at the time of writing of \cite{walkeretal2017}, less than three months after discovering IDV in \source, its annual cycle was not established and the kinematics of the screen were unknown. 

In this paper we report the results of tracking the rate of flux density variations in \source for just over a year, from February 2017 to February 2018, \upd{in which an evolution in the scintillation timescale is clearly seen. We interpret this evolution in purely kinematic terms -- i.e. we assume that it is an annual cycle arising from the Earth's orbit -- but with only one year of data we are unable to demonstrate the repetition that is expected for an annual cycle. In Section~\ref{section:data} we present the observations and data reduction. Section~\ref{section:inference} describes our inference of variability rates;} our method allows us to characterise the scintillation rate during slow phases of the cycle, where traditional methods of analysis struggle. Section~\ref{section:kinematics} fits the kinematic parameters of the annual cycle to the data and compares the results to the predictions of the model that connects the scattering with Spica. We discuss the degeneracy in our constraints Section~\ref{section:discussion} before concluding in Section~\ref{section:conclusions}.

\section{Observations and Data reduction}
\label{section:data}
We observed \source with the Australia Telescope Compact Array (ATCA) at 18 epochs, taking between 30 and 110 spectra extending from approximately $4.3\,\mathrm{GHz}$ to $11\,\mathrm{GHz}$ using two quasi-simultaneous tunings, of $150\,\mathrm{s}$ integration time each. The summary of these data is given in Table~\ref{table:epochs}. 

\begin{table}
	\centering
	\caption{Parameters of the 18 observational epochs on which long light curves were obtained. The right column shows the number of data points remaining in the $(5.5\pm0.25)/(10\pm0.25)\,\mathrm{GHz}$ sub-bands, as used in Section~\ref{section:inference}, after RFI excision.}
	\label{table:epochs}
	\begin{tabular}{lcrrr} 
		\hline
		Epoch & Date & D.o.Y. & MJD (mean) & \#points\\
		\hline
		1 & 2017/02/02 & 32 & 57785.70 & 104/103\\
		2 & 2017/02/07 & 37 & 57790.69 & 75 \\
		3 & 2017/02/19 & 49 & 57802.77 & 30 \\
		4 & 2017/02/21 & 51 & 57804.72 & 73 \\
		5 & 2017/02/23 & 53 & 57806.68 & 90 \\
        6 & 2017/03/23 & 81 & 57834.65 & 105/104\\
        7 & 2017/03/24 & 82 & 57835.65 & 102/103\\
        8 & 2017/04/11 & 100 &57853.67 & 75 \\
        9 & 2017/05/10 & 129 &57882.52 & 105\\
        10& 2017/05/11 & 130 &57883.51 & 109\\
        11& 2017/08/14 & 226 &58979.25 & 39 \\
        12& 2017/08/30 & 242 &58995.15 & 70 \\
        13& 2017/10/01 & 274 &58027.10 & 109/108\\
        14& 2017/10/02 & 275 &58028.11 & 0/0\\
        15& 2017/11/02 & 306 &58059.02 & 105/103\\
        16& 2017/12/15 & 348 &58101.90 & 107\\
        17& 2017/12/16 & 349 &58102.90 & 90\\
        18& 2018/02/24 & 54(+365) &58172.72 & 46\\
		\hline
	\end{tabular}
\end{table}

\begin{figure*}
	\includegraphics[width=0.75\linewidth]{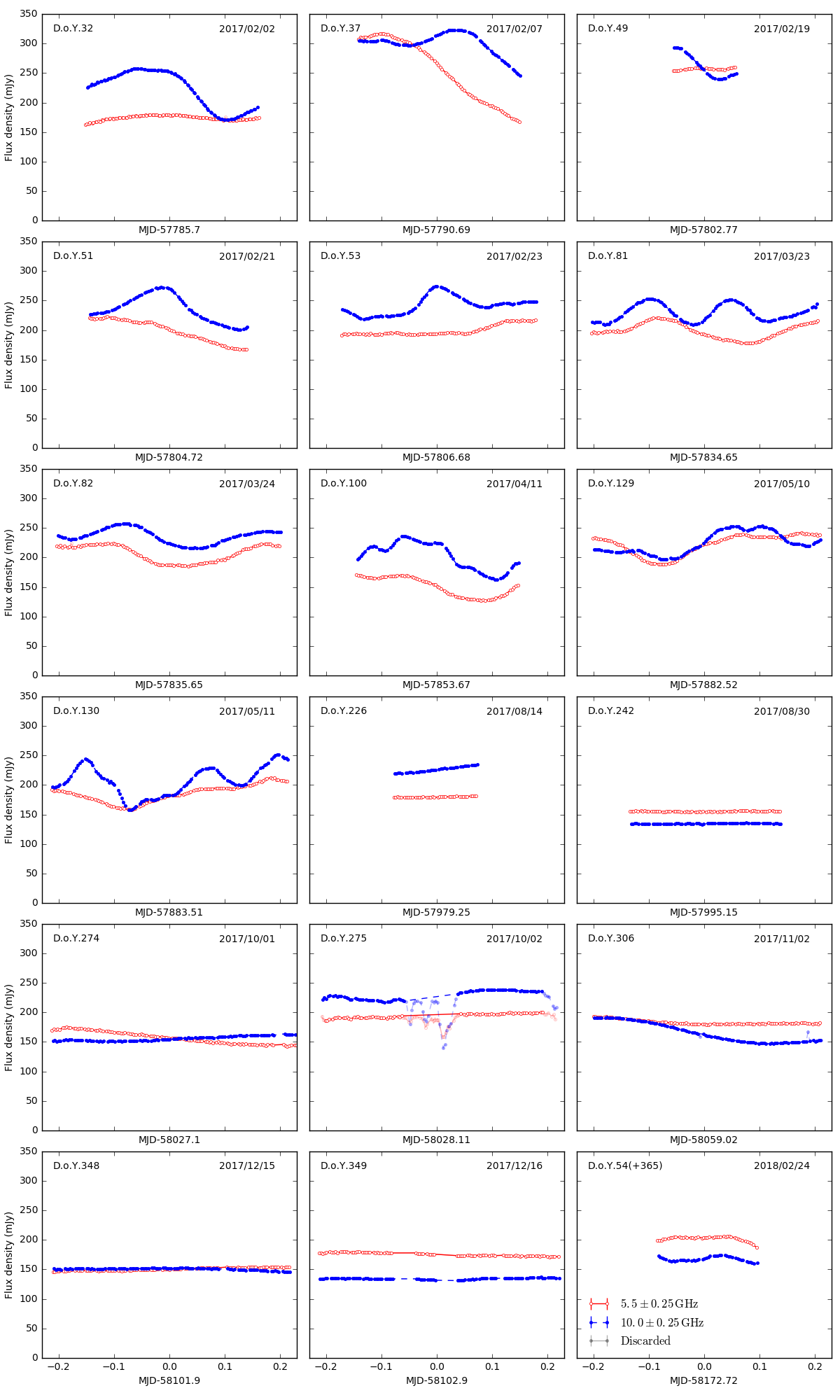}
    \caption{Observed light curves of \source, averaged over $0.5\,\mathrm{GHz}$-wide bands near the bottom and top of our bandwidth. The error bars, mostly too small to be seen, estimate the uncertainty of the mean. The points shown in semi-transparent were discarded from the analysis.}
    \label{figure:lightcurves}
\end{figure*}

To form the light curves used in the variability rate analysis below, we first filtered outliers from each recorded spectrum in the sub-bands of interest, by discarding data points that deviated from the mean of the group of their 10 closest neighbours by more than 3 times the r.m.s. values of the group, repeating this procedure twice on the updated spectra. We then visually inspected the full dynamic spectra and dropped those remaining data points that were clearly affected by RFI or other instrumental issues. In particular, we have excluded the entire data set from 2017/10/02 due to a persistent low-amplitude `moire' pattern in the dynamic spectrum; the origin of this pattern is unclear. Figure~\ref{figure:lightcurves} presents the light curves of \source observed at all 18 epochs, averaged over two $0.5\,\mathrm{GHz}$-wide bands centred at $5.5\,\mathrm{GHz}$ and $10\,\mathrm{GHz}$. 

\section{Variability rate inference}
\label{section:inference}

Traditionally, IDV have been analysed by direct computation of the temporal auto-correlation function (ACF) \citep[e.g.,][]{cycle1257}, or the structure function \citep[e.g.,][]{cycle1128but}, from the measured flux densities; and then extracting a characteristic timescale. However, either of these methods is difficult to use during slow phases in the cycle when the entire observing run might not be long enough to record a single oscillation in the light curve, leading to a biased estimate of the ACF. The epochs of slow variation are nevertheless particularly useful in constraining the kinematics of the screen, as we will see in \S4.3.2, and we are therefore motivated to try a new approach to estimating the variability timescale.

We model the light curves as a realisation of a stationary Gaussian process with an epoch-dependent time axis scaling; \upd{this is an assumption that is motivated by simplicity and the availability of suitable tools for the subsequent analysis.} Let $f_{ij}$ be the $j$-th data point of the $i$-th light curve measured at time $t_{ij}$ with some uncertainty $\sigma_{ij}$ uncorrelated between the measurements. Neglecting the correlations between any of our $I$ epochs and assuming that the variations on a given epoch are due to the time-dependent magnification of the intrinsic flux density $\overline{f}_i$, the likelihood of observing these data for a Gaussian process is given by the product
\bea\label{likelihood}
p\left(\{f_{ij}\}, \{t_{ij}\}\right)=\prod\limits_{i=1}^I{\det}^{-1/2}\left(2\pi\rm{M}_i\right)\times\hspace{2.7cm}\\\nonumber\exp\left[-\frac12\sum\limits_{jj'}\left(f_{ij}-\overline{f}_i\right)\left(\rm{M}_{i}\right)^{-1}_{jj'}\left(f_{ij'}-\overline{f}_i\right)\right],
\eea
where $\rm{M}_i$ is the covariance matrix given by the sum of the uncorrelated measurement noise and the matrix of the autocorrelation function (ACF) values at the observed time lags
\bea\label{covariancematrix}
M_{i\, jj'}=\delta_{jj'}\sigma^2_{ij}+\overline{f}_i^2 K_i\left(t_{ij}-t_{ij'}\right).
\eea
The auto-correlation functions $K_i(\Delta t)$ are unknown but we will assume that they derive from the same underlying $K(\Delta t)$ by an epoch-dependent rescaling of its argument:
\bea\label{kernelscaling}
K_i(\Delta t)=K(r_i \Delta t), \hspace{.5cm}i=\overline{1,I}.
\eea
That assumption reflects the expectation that the variations are statistically uniform in space, with the rate varying only because of the changing orbital velocity of the Earth.  The likelihood~(\ref{likelihood}) thus encodes information on the rates, $r_i$, via the Bayes theorem.

At the outset of our study it was unclear whether or not a Gaussian process should provide a good description of the IDV phenomenon, and our choice of model was driven mainly by the need for a method that is both tractable and unbiased when the variability timescales are long. We will see later that our approach has proved only partially successful.

In practice, the most computationally expensive part of evaluating~(\ref{likelihood}) is calculating the determinant of $\rm{M}$ and the value of the quadratic form in the argument of the exponent, given $K_i$. \upd{Unless the kernel is chosen very carefully, with numerical efficiency in mind, Bayesian inference is not computationally feasible even for moderately-sized datasets. Recently} a fast algorithm, {\tt celerite}, was developed (\citealt{celerite}, see also \citealt{rybickipress1995}) that can be used to compute the likelihoods~(\ref{likelihood}) very efficiently for a class of kernels $K_i(\Delta t)$ represented by a sum of exponentials with complex coefficients, of which we only consider even functions:
\bea\label{celeritekernel}
K(\Delta t)=\sum\limits_{n=1}^N a_n e^{-c_n\Delta t}  \cos d_n \Delta t, \hspace{.5cm} a_n, c_n, d_n>0.
\eea
\upd{This form appears well matched to the damped oscillations seen in the correlation functions of published IDV light-curves \citep[e.g][]{cycle1257}.}
We make use of the algorithm realised as a Python package, {\tt celerite}, released with the \citet{celerite} paper along with the {\tt emcee} package \citep{emcee} for Markov Chain Monte Carlo (MCMC) implementation. We experimented with the number of terms in~(\ref{celeritekernel}) by running the optimisation code {\tt NLopt} (Johnson 2010\footnote{http://ab-initio.mit.edu/nlopt}, Powell 2009\footnote{http://www.damtp.cam.ac.uk/user/na/NA\_papers/NA2009\_06.pdf}) for a fixed number of objective function evaluations and various $N$ and compared them using the adjusted Akaike information criterion \upd{\cite[e.g.][]{maier2013}}, which produced more consistent results compared to similarly used Bayesian information criteria. In most cases the optimum value turned out to be $N=1$ and it was never above four; moreover, the results for $r_i$ did not seem to be much affected if just a single term was used. We thus settled on $N=1$ in the MCMC calculations.

Rather than a single light curve, we record thousands of spectral channels, which all contain information on the rate of flux density variation. However, neighbouring channels are not independent with a decorrelation scale of a few GHz on most epochs, and the likelihood of the entire data set would need to account for correlations along the frequency axis \upd{by adding a pair of indices in addition to $j,j'$ in~(\ref{likelihood}) -- and thus greatly increasing the dimensionality of the parameter space}. This correlation structure is of little interest for the present work but adds substantial computational expense. To use as much data as possible and at the same time keep calculations practically feasible \upd{we extract two light curves by frequency-averaging the data near the edges of the observed bandwidth, one centred at $5.5\,\mathrm{GHz}$, the other at $10\,\mathrm{GHz}$,}\upd{ and replace~(\ref{likelihood}) with a product of two such expressions, one for each sub-band (which assumes that the two light curves are not correlated)}. \upd{We use $0.5\,\mathrm{GHz}$-wide intervals}, the width where the empirical uncertainty of the mean over the interval (which includes both noise and real variations with frequency) approaches the expected thermal noise in the interval. This value is $\sim0.3\,\mathrm{mJy}$ for both sub-bands, and we use the empirical uncertainty of the mean as a measure of $\sigma_{ij}$ in~(\ref{covariancematrix}). We use the arithmetic mean of the light curve as an estimate of the intrinsic flux density, $\overline{f}_i=\langle f_{ij}\rangle_j$, allowing for its variation from epoch to epoch as appropriate for a compact flat spectrum source. We attempted to explore the $\overline{f}_i$ parameter space with the MCMC method but failed to achieve convergence even after running the chains for several days -- presumably due to the significant dimensionality added. The factors $r_i$ are the same for both light curves. As the likelihood is invariant to scaling of all rates by the same factor while simultaneously scaling $c_n, d_n$ coefficients by its inverse, a reference rate is specified by keeping $ r_1\equiv1$ in the code. There are therefore $I-1+2\cdot 3$ parameters \upd{(rates for $I-1$ epochs relative to the first epoch plus 3 model ACF parameters for each of 2 sub-bands)} which are all positive and assumed {\it a~priori} distributed uniformly in log between $e^{-15}$ and $e^{15}$, measured in $\mathrm{d}^{-1}$ for $c_n, d_n$, and dimensionless otherwise. 

\begin{figure}
	\includegraphics[width=\columnwidth]{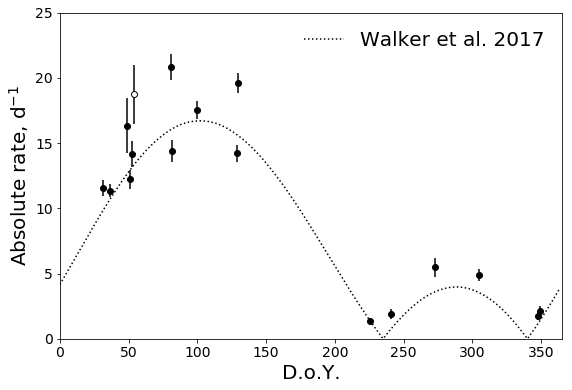}
    \caption{Absolute variation rates ($R_i$, from equation \ref{absoluteratedef}), versus day of the year, at $5.5\,\mathrm{GHz}$ (the $10\,\mathrm{GHz}$ sub-band behaves very similarly\upd{-- by construction, they only differ in the normalisation, as relative rates are the same for both sub-bands}). The open symbol shows our single epoch of observation in 2018. Error bars \upd{extend over 68 per cent variation of the MCMC samples}; points mark the medians of the distributions. The dotted line shows the prediction of a highly anisotropic model that associates the scattering plasma with Spica \citep{walkeretal2017}\upd{ -- specifically, assumes that the plasma co-moves with the star while its anisotropy axis points at it and the source has a brightness temperature of $T_b=10^{13}\,\mathrm{K}$}.}
    \label{figure:rates}
\end{figure}

Figure~\ref{figure:rates} presents the inferred 
absolute rates, $R_i$, defined as the inverse of the ACF half-width at half-maximum (HWHM):
\bea\label{absoluteratedef}
R_i=\frac1{\tau_i}=\frac{r_i}{\tau}, \hspace{.5cm}\mathrm{where}\,\tau: K(\tau)=\frac{K(0)}2.
\eea
We choose to plot the scintillation rates rather than timescales because, for a one-dimensional model, the former are proportional to a component of the effective velocity, $v^\perp_\mathrm{eff}$. As such the rates are expected to be a sinusoidal function of time, and the information content of the data is readily perceived. In \S4.3 we give a qualitative analysis of the kinematic constraints that can be obtained from our data; for now we note that the phase of the sinusoid reflects the orientation of the plasma anisotropy, and the offset reflects the corresponding component of the plasma velocity.

\section{Fitting the annual cycle}
\label{section:kinematics}

\subsection{Performance of the Walker et al. 2017 model}
\label{subsection:walkeretal model}

The dotted line in Figure~\ref{figure:rates} shows the prediction of the \cite{walkeretal2017} model, in which the scattering plasma is co-moving with Spica and highly anisotropic with its major axis pointing at the star; it has no free parameters. The absolute rate of the variations,
\bea\label{absoluterate}
R=\frac{v^\perp_\mathrm{eff}}{a_\perp},
\eea
is determined by the effective transverse velocity \citep{effectivevelocity},
\bea\label{veff}
{\bf v}_\mathrm{eff}={\bf v}_\mathrm{screen}-{\bf v}_\oplus,
\eea
and $a_\perp$, the HWHM of the ACF of the spatial structure of the scintillation pattern.

Qualitatively, the model prediction appears broadly consistent with the shape of the inferred variation rate, which depends on the kinematics only, but appears slightly off in normalisation -- i.e., the model value of $a_\perp$ is too high. \cite{walkeretal2017} associate the latter with the angular width of the source projected through the screen, at distance~$D_s$, into the observer's plane, assuming a Gaussian source of peak brightness temperature $T_b=10^{13}\,\mathrm{K}$. The ACF HWHM expected at the wavelength $\lambda$ is then
\bea\label{sourcehwhm}
a_\perp=\lambda  D_s\sqrt{\frac{\overline{f} \log 2}{\pi k_B T_b}}
\eea
and Figure~\ref{figure:rates} uses an average $\overline{f}_i$ of $191\,\mathrm{mJy}$, resulting in $a_\perp\approx2.26\times10^{5}\,\mathrm{km}$ for $5.5\,\mathrm{GHz}$ and (Spica) distance of $77\,\mathrm{pc}$. In reality $a_\perp$ is also influenced by other aspects of the problem --- the Fresnel scale and the strength of the scattering, for example \citep{GoodmanNarayan2006}. However, apart from normalisation, the model appears to be consistent with the positions of standstills as well as both the positions and relative amplitudes of the two peaks of the rate annual curve.

In quantitative terms, the model presented by the dotted line in Figure~\ref{figure:rates} is a poor fit to the data; its $\chi^2$ computed from the numbers in the figure is $101.4$\upd{. As the fitting procedure treats the 17 rates inferred in Section~\ref{section:inference} as effective measurements (with associated uncertainties) and the model it is compared to has not been fitted to this data, the expected value of $\chi^2$ is 17, much lower than measured.} Treating $a_\perp$ (estimated to be uncertain by $\sim 0.5\,\mathrm{dex}$ in \citealt{walkeretal2017}), as an additional free parameter reduces the discrepancy measure to 83.5, still much above the expectation \upd{of $16=17-1$}. Interestingly though, by varying all \upd{three }parameters of a one-dimensional model \upd{-- $a_\perp$, orientation of the plasma anisotropy axis and plasma velocity relative to this axis, {\it c.f.}~(\ref{veffperp}) and immediately thereafter -- }the best fit one can achieve has a $\chi^2$ of 75.3, still far above the expected value of \upd{$14=17-1-2$}. Although various interpretations are possible, this comparison suggests that the error bars in Figure~\ref{figure:rates} are underestimated because our statistical model is not entirely adequate.

That would not be surprising, given that the Gaussian process assumption and the chosen ACF parameterisation were motivated largely by a need for feasible computations\footnote{\upd{Unfortunately we are at present unable to suggest any better statistical models.}}. Although any reasonable choice of the likelihood functional form would push the model to some sort of match with the data, MCMC methods use the likelihood ratios to decide how often a certain region of the parameter space is to be explored, and the convergence dynamics may be adversely affected by a poor statistical description. The foregoing concerns about the statistical model are reinforced by examination of our MCMC determinations of the scintillation rates for the individual epochs. In Figure~\ref{figure:rates} we can see examples where consecutive days -- $\mathrm{D.o.Y.}\,81-82$ and $\mathrm{D.o.Y.}\,129-130$ -- yield highly significant differences between the inferred rates, despite qualitatively similar light curves (Figure~\ref{figure:lightcurves}). By contrast, no significant difference is expected between consecutive days if the evolution of the rates is purely of kinematic origin, as the Earth's orbital velocity changes only slightly from day to day. We also note that epochs close to the expected standstills, where very low rates are inferred, yield small fractional rate uncertainties, whereas we expect fractional uncertainty of order unity, because of sample variance (less than one oscillation sampled within our observing window). Rescaling the error bars by the square-root of the $75.3:14$ ratio of $\chi^2$ values (actual:expected) so as to bring the best-fit model reduced $\chi^2$ to unity makes the \cite{walkeretal2017} model consistent with the data at better than $1\,\sigma$ level. Figure~\ref{figure:fitrates} illustrates the effect of rescaling on consistency of rate estimates on consecutive days.

\subsection{Kinematic MCMC fitting}
\label{subsection:kinematicsmcmc}

The least squares analysis of the kind described in the previous subsection does not take into account correlations between the rate estimates, which arise due to the dependence of the likelihood~(\ref{likelihood}) on the global parameters of the ACF model as well as the individual $r_i$. It is therefore most appropriate to fit the kinematic model directly to the light curves by substituting $r_i=R(t_i)/R(t_1)$ with absolute rates $R_i$ given by~(\ref{absoluterate}) and running the MCMC code on the parameter space of the kinematic model. 

For a highly anisotropic plasma screen, the expected variation rate is proportional to the (absolute value of the) component of transverse effective velocity 
across the anisotropy axis,
\bea\label{veffperp}
a_\perp R_i=\left|{\bf v}_\mathrm{eff}(t_i)\hat{\bf e}^\perp\right|=\left|v_\mathrm{screen}^\perp-v_\oplus(t_i)\sin\left[\mathrm{PA}_\oplus(t_i)-\mathrm{PA}\right]\right|.
\eea
The parameter space of our kinematic model is thus two-dimensional, spanned by the position angle of the anisotropy axis, $\mathrm{PA}$, and the component of the screen velocity orthogonal to that axis, $v_\mathrm{screen}^\perp$. The velocity component parallel to the anisotropy axis does not affect rates for the one-dimensional model and cannot be constrained. It is convenient to keep $v^\perp_\mathrm{screen}$ non-negative with $\mathrm{PA}\in[0,2\pi)$ such that 
\bea
\hat{\bf e}^\perp\equiv\hat{\bf e}_\mathrm{Dec}\cos\left(\mathrm{PA}+\frac\pi2\right)+\hat{\bf e}_\mathrm{RA}\sin\left(\mathrm{PA}+\frac\pi2\right)
\eea
and ${\bf v}_\mathrm{screen}$ make an acute angle:
\bea\label{paconvention}
v_\mathrm{screen}^\perp\equiv\left(\hat{\bf e}^\perp\cdot{\bf v}_\mathrm{screen}\right)\geq0.
\eea
We calculate the Earth barycentric velocity transverse to the line of sight using the {\tt get\_body\_barycentric()} function of the {\tt astropy.coordinates} package \citep{astropy2018}. To handle the non-trivial topology of the $\mathrm{PA}$ domain in the MCMC exploration, technically we reparametrise the problem 
\bea\label{xy}
v_\mathrm{screen}^\perp, \mathrm{PA}\,\to\, x, y: \left(\begin{array}{l}x\cr y\end{array}\right)=v_\mathrm{screen}^\perp\left(\begin{array}{l}\cos\mathrm{PA}\cr\sin\mathrm{PA}\end{array}\right)
\eea
The priors on $x,y$ are flat within $\pm300\,\mathrm{km}\,\mathrm{s}^{-1}$. The MCMC implementation is otherwise the same as for the rate estimates.

\begin{figure}
	\includegraphics[width=\columnwidth]{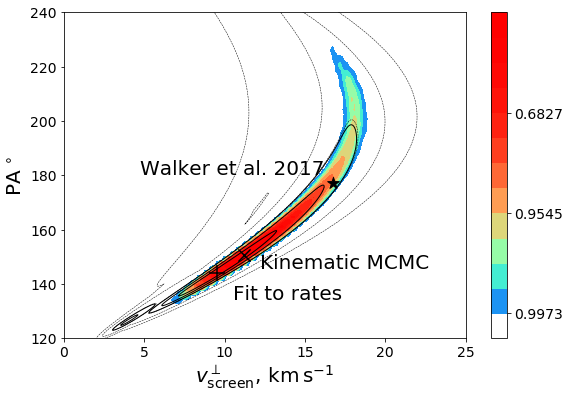}
    \caption{Posterior distribution of the parameters of the anisotropic scattering screen. Colours represent the confidence intervals of the MCMC posterior distribution\upd{, as described in \S\ref{subsection:kinematicsmcmc},} around its densest point, marked with a cross. Contour lines show the results of $\chi^2$ fitting to the rate estimates shown in Figure~\ref{figure:rates} (ignoring correlations between those estimates)\upd{, as described in \S\ref{subsection:walkeretal model}}. Contours are drawn at the levels marked in the colour bar, which correspond to the standard 1-, 2- and 3-$\sigma$ levels of a 1D Gaussian distribution. Solid contours show $\chi^2$ computed using the rates with unscaled error bars (as shown in Figure~\ref{figure:rates}); dotted contours show $\chi^2$ computed using the rates with error bars scaled up by a factor of 2.32 (as shown in Figure~\ref{figure:fitrates}), which correspond to a minimum $\chi^2$ of 14. The location of the minimum is marked with a plus. The star corresponds to the model of \citet{walkeretal2017} that attributes scattering to radial plasma filaments co-moving with Spica; the model is near the 2-$\sigma$ contour of the MCMC and unscaled $\chi^2$ analysis, and not significantly different from the best $\chi^2$ fit if error bars are inflated.}
    \label{figure:kinematics}
\end{figure}

Figure~\ref{figure:kinematics} presents the results of this modelling, reparameterised back to the $(v_\mathrm{screen}^\perp, \mathrm{PA})$ space. The maximum density of the samples, represented as shades, is formally achieved near $(v_\mathrm{screen}^\perp, \mathrm{PA})\approx(11.2\,\mathrm{km}\,\mathrm{s}^{-1}, 151^\circ)$ but an extended region of the parameter space is consistent with the data. In particular, the position of the \cite{walkeretal2017} model, $(v_\mathrm{screen}^\perp, \mathrm{PA})\approx(16.8\,\mathrm{km}\,\mathrm{s}^{-1}, 177^\circ)$, is consistent with the data at just inside the $2\sigma$ level. Also marked at $(v_\mathrm{screen}^\perp, \mathrm{PA})\approx(9.5\,\mathrm{km}\,\mathrm{s}^{-1}, 144^\circ)$ is the model that minimises the $\chi^2$ difference between the kinematic model and the rates as shown in Figure~\ref{figure:rates}. Solid lines show the confidence intervals of this $\chi^2$ statistic whereas light dotted lines are the same levels after rescaling the error bars so as to bring the reduced $\chi^2$ of the best fit to unity.

A word of caution concerning the kinematic fitting is that our Markov Chains could not achieve convergence, as judged by the conventional heuristics \citep{hoggforemanmackey2017}. 
However, we do not observe any indications of unusual behaviour of the {\tt emcee} walkers, nor do we see significant multi-modality of the posterior distribution. Why the autocorrelation time estimates continue to rise almost linearly with the chain length is not clear; one possibility is an inadequacy of the statistical model, as discussed above. As a result, we are not certain which set of lines or shades in Figure~\ref{figure:kinematics} best represents our uncertainty about the kinematic parameters of the scattering plasma, but it is clear that this region extends over a large range of position angles.

\begin{figure}
	\includegraphics[width=\columnwidth]{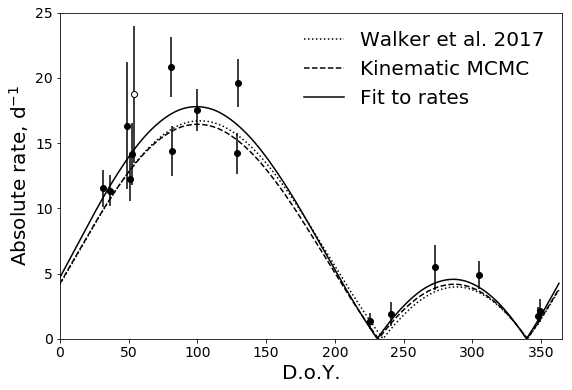}
    \caption{Annual cycles of the variation rate for the three models marked in Figure~\ref{figure:kinematics} compared to the values inferred in Section~\ref{section:inference} (with inflated error bars). Despite considerable difference in the input parameters, particularly in the orientation of the anisotropy, the differences in the predicted curves are small and, for these error bars, not significant.}
    \label{figure:fitrates}
\end{figure}

Figure~\ref{figure:fitrates} partly explains the significant extent of this region by showing the performance of the three models pinpointed in Figure~\ref{figure:kinematics} in fitting the variation rates inferred from the light curves. Qualitatively, the three model curves perform similarly well and appear hardly distinguishable given the quality of constraints presented.

\subsection{Qualitative analysis}
\label{subsection:kinematicsqualitative}

We will now explain the origin of the degeneracy in the kinematic parameters seen in Figure~\ref{figure:kinematics}, see what inferences can be drawn directly from the rate curve and describe how the annual cycle can be analysed qualitatively.

\subsubsection{Hodograph}
\label{subsubsection:hodograph}

\begin{figure}
	\includegraphics[width=\columnwidth]{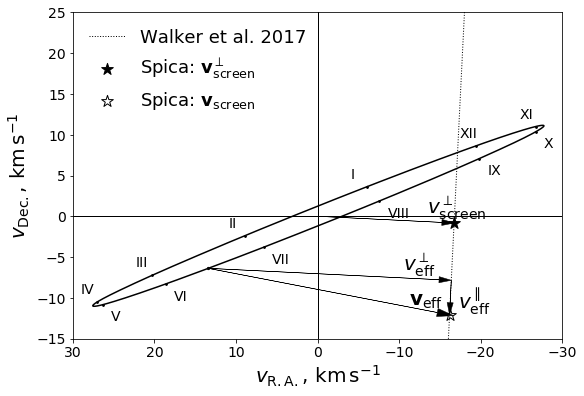}
	\includegraphics[width=\columnwidth]{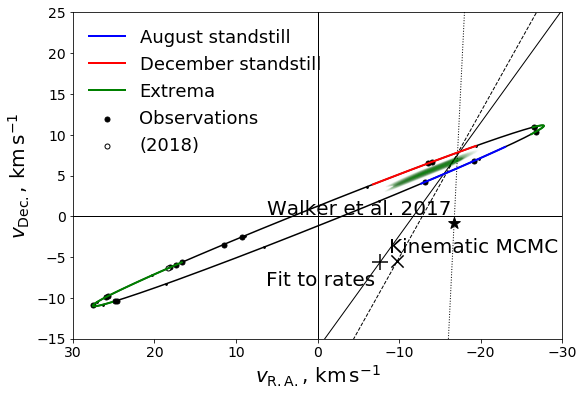}
    \caption{{\it (Top)} Hodograph of the Earth orbital velocity transverse to the direction to \source. Roman numerals mark the beginning of respective months. The effective velocity ${\bf v}_\mathrm{eff}$ is given by the vector from the point on the hodograph to the transverse velocity of the screen (shown with an open star for Spica). In highly anisotropic scattering, the scintillation rate depends on the component  $v^\perp_\mathrm{eff}$ orthogonal to the anisotropy axis (shown with a dotted line for \source-Spica orientation). {\it(Bottom)} The hodograph displaying the location of standstills and extrema of the rate from Figure~\ref{figure:rates} along with the observation dates. The successful kinematic model should pass through the uncertainty intervals of both standstills on the hodograph as well as a shaded region inside the curve, representing the constraint~(\ref{extremahodograph}) from the extrema. Also shown are the three kinematic models marked in Figure~\ref{figure:kinematics}; all do a reasonable job reproducing standstills but struggle somewhat to fit the observed extremal rates.}
    \label{figure:hodograph}
\end{figure}

In this section we will extensively use the hodograph of the velocity vector, which is the locus of the terminal points of a variable vector as its initial point is held fixed. In particular, the top panel of Figure~\ref{figure:hodograph} plots the hodograph of the component ${\bf v}_\oplus$ of the Earth velocity that is transverse to the line of sight to \source. Roman numerals along the curve mark the start of the corresponding months. The hodograph is, to a high accuracy, an ellipse, 
and the low ecliptic latitude of \source (it is just $2^\circ$ below the ecliptic plane) gives the hodograph its highly elongated shape. 

According to~(\ref{veff}), the effective velocity ${\bf v}_\mathrm{eff}$ on a particular date is given by the vector from the corresponding point on the hodograph curve to a fixed point in the plot, the transverse velocity of the screen, ${\bf v}_\mathrm{screen}$. As an example, Figure~\ref{figure:hodograph} shows the transverse velocity of Spica with an open star symbol. For highly anisotropic scattering, only one component of the effective velocity affects the rate of scintillation, the component $v_\mathrm{eff}^\perp$ orthogonal to the major axis of the illumination pattern, at a position angle of $\mathrm{PA}$. Geometrically, this component is equal to the distance between the respective point on the hodograph curve and the straight line that passes through ${\bf v}_\mathrm{screen}$ and has the position angle of the pattern anisotropy axis. In contrast, the effective velocity component along this line has no bearing on the scintillation timescales; conversely, this component cannot be deduced from the annual cycle analysis. Hence, the screen transverse velocity implied by the annual cycle can lie anywhere along this line and therefore the line itself represents the kinematic model:
\bea\label{kinematicline}
{\bf v}_\mathrm{screen}\cdot\hat{\bf e}^\perp=v_\mathrm{screen}^\perp.
\eea
Its two parameters are the position angle $\mathrm{PA}$ and distance from the origin, $v_\mathrm{screen}^\perp$, with an $n\times\pi$ ambiguity in the direction of the major axis resolved by~(\ref{paconvention}) such that $v_\mathrm{screen}^\perp\geq0$. Line representation is useful for qualitative analysis of the most salient features of the rate curve -- the position of its standstills (where present), and the positions and relative magnitudes of the rate local maxima. They have a very simple interpretation on the hodograph plot: the standstills are the intersections of the hodograph with the model line (so that the rate, proportional to $v_\mathrm{eff}^\perp$ is zero), and at the extrema the hodograph tangents are parallel to the line (so that the line-hodograph distance is stationary). The bottom panel of Figure~\ref{figure:hodograph} shows lines representing the three kinematic models identified in Figure~\ref{figure:kinematics}.

An alternative -- and more compact -- representation of the kinematic model is by the end point of the vector
\bea
{\bf v}^\perp_\mathrm{screen}\equiv v^\perp_\mathrm{screen}\hat{\bf e}^\perp
\eea
(which is $\not\equiv{\bf v}_\mathrm{screen}$, and neither to $ v^\perp_\mathrm{screen}$ as it also depends on the orientation of the anisotropy axis); a filled star in Figure~\ref{figure:hodograph} marks ${\bf v}^\perp_\mathrm{screen}$ of the \cite{walkeretal2017} model. While less natural for qualitative analysis, this representation is convenient when considering multiple models simultaneously -- e.g., when comparing them or representing uncertainty regions, which are difficult to visualise unambiguously for sets of straight lines. 

\subsubsection{Standstills}

If the annual curve contains standstills, where the rate drops to zero, accurate knowledge of their positions is sufficient to determine $\mathrm{PA}$ and $v_\mathrm{screen}^\perp$ with no input from variable epochs. Since the rate is given by the distance from the point on the hodograph to the straight line representing the kinematic model $(\mathrm{PA}, v_\mathrm{screen}^\perp)$, this distance should be zero at standstill -- {\it i.e.,} the line should pass through the standstill. This is also true of the second standstill, and the two therefore completely define the kinematic model. Formally, by requiring~(\ref{veffperp}) to vanish, one obtains a sinusoid in the $(v_\mathrm{screen}^\perp, \mathrm{PA})$ plane,
\bea\label{standstill}
v_\mathrm{screen}^\perp=v^\mathrm{stand}_\oplus\sin\left(\mathrm{PA}^\mathrm{stand}_\oplus-\mathrm{PA}\right),
\eea
where 
$v^\mathrm{stand}_\oplus$, $\mathrm{PA}^\mathrm{stand}_\oplus$ are the magnitude and position angle of the Earth transverse velocity at the time of the standstill, both known. In the ${\bf v}^\perp_\mathrm{screen}$ representation, the condition is
\bea\label{standstillvector}
{\bf v}^\mathrm{stand}_\oplus\cdot\hat{\bf e}^\perp=v^\perp_\mathrm{screen}\hspace{.3cm} \Leftrightarrow\hspace{.3cm} \left({\bf v}^\mathrm{stand}_\oplus\cdot\hat{\bf e}^\perp\right)\hat{\bf e}^\perp={\bf v}^\perp_\mathrm{screen}
\eea
-- {\it i.e.}, that ${\bf v}^\perp_\mathrm{screen}$ is an orthogonal projection of ${\bf v}^\mathrm{stand}_\oplus$; the locus of such ${\bf v}^\perp_\mathrm{screen}$ is a circle of which ${\bf v}^\mathrm{stand}_\oplus$ is a diameter. The full  solution is obtained by locating the intersection of the two standstill sinusoids $(v^\perp_\mathrm{screen}, \mathrm{PA})$ or circles (${\bf v}^\perp_\mathrm{screen}$).

In practice, the position of a standstill is known with some uncertainty due to gaps in variability monitoring. In fact, it is not possible to know for sure if an observed lull in fluctuations is due to the Earth being stationary with respect to the structure in the flux distribution, or because the flux distribution happens to have a locally flat area there. This replaces a pair of points on the hodograph track with a pair of uncertainty regions that contain the standstills, and any line passing through both regions is a potential solution. Likewise, the constraint lines in the parameter planes are replaced by sinusoidal $(v^\perp_\mathrm{screen}, \mathrm{PA})$ or circular (${\bf v}^\perp_\mathrm{screen}$) bands of non-zero (and varying) width. 

Generally, the further apart the two uncertainty regions in the hodograph, the better constrained the position angle and the orthogonal component of the screen velocity are.
However, the case in Figure~\ref{figure:hodograph} seems to be closer to the other extreme, with the kinematic parameters quite uncertain. Looking at the light curves in Figure~\ref{figure:lightcurves} and inferred rates in Figure~\ref{figure:rates} one might argue that the standstills are observed around $\mathrm{D.o.Y.}\,240\pm15$ (in particular, the light curve on D.o.Y.~242 seems featureless in both sub-bands) and $\mathrm{D.o.Y.}\,350\pm15$, as highlighted in the bottom panel of Figure~\ref{figure:hodograph}. Because of the low ecliptic latitude of \source the hodograph is tightly squeezed in the latitudinal direction and the two uncertainty regions are quite close to each other. As a result, the position angle of the line passing through these constraints can be anywhere from $\sim 120^\circ$ to $290^\circ$ ($\mathrm{N}\to\mathrm{E}$) and its distance from the origin similarly varies from $\simeq3\,\mathrm{km}\,\mathrm{s}^{-1}$ to $\simeq22\,\mathrm{km}\,\mathrm{s}^{-1}$. 

We note that the configuration of the standstill seasons observed in \source is close to as bad as it gets in this sense. A different arrangement could, in principle, have been much better constraining; but in practice one is less likely to obtain a favourable configuration for a source that has low ecliptic latitude. One can typically expect much better constraints from standstills in the annual cycles of IDVs that are far from the ecliptic plane.

\subsubsection{Extrema}

Another easily interpretable trait of rate cycles, whether displaying standstills or not, is the timing and relative magnitude of the  local extrema. For a one-dimensional model, the tangents to the hodograph at the extrema positions are parallel to the anisotropy axis; to a very high accuracy, they should be opposite to each other in the hodograph (and very close to half a year apart in the rate curve). The relative magnitudes of the rate extrema in turn require the kinematic constraint line to pass through a point ${\bf v}_r$ on the line connecting extrema positions ${\bf v}^\mathrm{ext}_{\oplus, 1,2} $, whose distances to the two points are in the same ratio as the respective extremal rates, $r_{1,2}$. There are two such points, one in between the extrema positions and the other behind the extremum ${\bf v}^\mathrm{ext}_{\oplus, 1}$ of smaller magnitude $r_1$:
\bea\label{extremahodograph}
{\bf v}_r={\bf v}^\mathrm{ext}_{\oplus, 1} (1-q)+{\bf v}^\mathrm{ext}_{\oplus, 2} q, \hspace{.5cm}
q=\pm\frac{1}{\frac{r_2}{r_1}\pm1}, \hspace{.2cm}r_1<r_2;
\eea
only the in-between case ($+$) is consistent with cycles displaying standstills for a one-dimensional model. 

The requirement that the constraint line passes through this point
\bea\label{extremacircle}
{\bf v}_r\cdot\hat{\bf e}^\perp=v^\perp_\mathrm{screen}
\eea
represents a circle of diameter ${\bf v}_r$ in the ${\bf v}_\mathrm{screen}^\perp$ space, or a sinusoid,
\bea\label{extrema}
v^\perp_\mathrm{screen}=v_r\sin\left(\mathrm{PA}_r-\mathrm{PA}\right),
\eea
in the $(\mathrm{PA}, v^\perp_\mathrm{screen})$ space. In the Solar system barycentre frame we expect ${\bf v}^\mathrm{ext}_{\oplus, 1}\approx-{\bf v}^\mathrm{ext}_{\oplus, 2}$ very accurately, hence
\bea
{\bf v}_r\approx{\bf v}^\mathrm{ext}_{\oplus,1}(1-2q)\hspace{.2cm} \Leftrightarrow\hspace{.2cm} v_r\approx v^\mathrm{ext}_{\oplus, 1}\frac{r_2\mp r_1}{r_2\pm r_1}, \hspace{.2cm}\mathrm{PA}_r\approx\mathrm{PA}^\mathrm{ext}_1;
\eea
the upper sign is to be taken for cycles with standstills.

Similarly to the standstill constraints, uncertainties in the measured positions and relative magnitudes of the extrema blur the location of the  point through which the kinematic constraint line should pass. Figure~\ref{figure:rates} suggests that the extrema are located around $\mathrm{D.o.Y.}\,290\pm20$ and $\mathrm{D.o.Y.}\,100_{-50}^{+30}$, with the latter particularly uncertain, and the ratio of the extremal rates is about $3_{-1}^{+2}$. The constraint that corresponds to these estimates is shown in Figure~\ref{figure:hodograph} as a shaded region inside the hodograph of the Earth's transverse velocity, and provides an upper limit on the magnitude of ${\bf v}_\mathrm{screen}^\perp$ as well as a sign of its projection on the ecliptic plane. On the other hand, the extrema themselves are located near the `pointy' ends of the hodograph where the position angle of the tangent line makes effectively a full swing rendering the inference of $\mathrm{PA}$ unreliable. Similarly to the case of standstills, this is partly due to the low ecliptic latitude of \source.

\vspace{.5cm}
\noindent The standstill and extrema constraints are not automatically consistent but they have to be so for a one-dimensional scattering model, which might help identify the limits of its applicability to the data. For more general anisotropy the situation can be quite different -- for example, standstills are not generally expected in such models, as that would require the two components of the effective velocity to vanish simultaneously. Likewise, the extrema are not generally expected to lie on the opposite sides of the hodograph, with a six month separation. The analysis of the more general case goes beyond the scope of this paper.

\section{Discussion}
\label{section:discussion}

As is clear from the qualitative analysis of the hodograph, accurate knowledge of two standstill positions is sufficient to uniquely determine the two kinematic parameters of the totally-anisotropic scattering model. However, near standstills the scintillation rate is difficult to constrain via traditional ACF estimation methods, and that was part of the motivation for our Bayesian analysis. For any cycle displaying standstills it is therefore highly desirable to obtain good coverage near the standstill epochs, to pin down the timing as tightly as possible. That is not practicable in the first monitoring season because the standstill positions are initially unknown, but becomes easier in subsequent years; we are currently obtaining such data for \source.

In this paper we have mostly concerned ourselves with the analysis of the shape of the annual cycle, while paying comparatively little attention to its absolute scale. The latter, however, is of interest in connection with the distance to the scattering screen. The prescription used by \cite{walkeretal2017} associates the spatial scale of the scintillation pattern with the source angular size multiplied by the screen distance. In the case of \source, and a screen placed at the distance of Spica, that prescription matched the value fit for the same $(v_\mathrm{screen}^\perp, \mathrm{PA})$ to better than $10$ per cent although, given the simplicity of the prescription -- \cite{walkeretal2017} suggested likely errors of $0.25$~dex -- this match must be considered a coincidence. One thing to bear in mind is that although specific assumptions about the character of the scattering plasma -- conventionally, a Gaussian random field with a power-law spectrum of density inhomogeneities -- allows one to calculate a precise value of the HWHM of the ACF of the scintillation pattern as a function of screen distance, pinning down the appropriate values of those parameters is not easy, and the unknown source structure (frequency-dependent size and shape) always remains a part of the mix. Indeed, the screen distance estimates obtained by such detailed modelling for the three best-studied IDVs (PKS~0405-385, J1819+3845 and PKS~1257-326) have  uncertainties of a factor of a few, and might vary by an order of magnitude between different models \citep{twostation1257,muasquasar,macquart2006,macquart2007,rkcj2002}. Furthermore, theoretical models often assume the scattering to be isotropic \citep[e.g.,][]{GoodmanNarayan2006} and are thus not applicable to the present data, for which the shape of the annual cycle indicates strong anisotropy; a similar resource for anisotropic plasma would be valuable.

\subsection{Does Spica host the scattering plasma?}
One of our main interests in following the scintillations of \source\ over the last year was the possibility of testing the suggestion of \citet{walkeretal2017} that the structures responsible for IDV are associated with foreground, hot stars. We have already shown (\S4) that our data are consistent with that model; but that could be a fortuitous agreement, arising in the context of a completely different model, and it is useful to evaluate the probability of such a coincidence. To do that we need to construct a statistical model for the distribution of the relevant parameters -- position angle and perpendicular velocity -- of the population of blobs of scattering plasma. Our adopted model is this: isotropic distribution of plasma microstructure orientation; and, an isotropic Gaussian distribution of the transverse velocity components. These properties are generic to a large class of models, and as such are reasonable assumptions. To fully specify the prior we need, in addition, a value of the dispersion of the velocity distribution. Rather than making an ad hoc assumption, we note that very small and very large values of the velocity dispersion would both yield vanishingly small probabilities of obtaining our results by chance. We therefore proceed by determining the value of the velocity dispersion that maximises the chances of a coincidence, so that the probability we obtain is an upper limit.

To compute that limit we integrate the probability distribution of our prior over the region of parameter space that matches our data at least as well as the \citet{walkeretal2017} prediction (Figure~\ref{figure:fitrates}).  Doing so we find a probability of 0.0051 (at a velocity dispersion of $13.0\,{\rm km\,s^{-1}}$), if the MCMC kinematic distribution is used to measure the quality of fit. If, on the other hand, we use the $\chi^2$ results then the probability is 0.0073 (at a velocity dispersion of $11.9\,{\rm km\,s^{-1}}$). Therefore we could indeed have obtained the observed agreement by chance, but the likelihood of doing so is less than 1\%. That figure is small enough to conclude that our data support the \citet{walkeretal2017} model, but not small enough to put it beyond doubt.

\section{Conclusions}
\label{section:conclusions}
Monitoring of PKS B1322$-$110 has revealed a strong annual cycle in the rate of its scintillations. The cycle, which appears to include two standstills, is consistent with a highly anisotropic model of the scattering plasma. Quantifying the timescale of the scintillations is challenging because it is often comparable to, and sometimes much larger than, the extent of the observing window. Using the {\tt celerite\/} MCMC package we evaluated the scintillation rate for each epoch, by determining a scaling factor for a temporal autocorrelation function whose shape is assumed constant over the year. The rates determined in this way exhibit highly significant differences between consecutive days, whereas the Earth's orbital velocity hardly changes at all, and we conclude that our statistical model underestimates the uncertainties. To bring the reduced $\chi^2$ of the best-fit model to unity we needed to increase the error bars on the rates by a factor of $2.3$.

Although our data for \source\ prefer a model in which the plasma microstructure points $\sim 30^\circ$ away from Spica, the preference is weak, and a radial orientation -- as suggested by \citet{walkeretal2017} -- is included in the 68\% confidence interval for the $\chi^2$ surface (after rescaling the error bars on the rates). The very broad angular distribution of the acceptable kinematic models is understood as a direct consequence of the the low ecliptic latitude ($-2^\circ$) of \source, which makes the shape of the annual cycle insensitive to microstructure orientation.

Despite the large uncertainty in the position angle of the plasma filaments, the relatively narrow range of preferred velocities at each orientation means that we have a one-dimensional constraint region in a two-dimensional space. Making use of a generic, isotropic prior for the statistics of the kinematics of the scattering material, we find a 1\%\ probability of discovering, by chance, a screen whose properties are at least as close to the data as those of the \citet{walkeretal2017} prediction. Our data thus provide some support for that model.

\section*{Acknowledgements}
\upd{We thank Bill Coles and J.-P. Macquart for insightful suggestions. AVT thanks Svetlana Nikolaevna Shlenova for introducing him to the concept of hodograph.} The Australia Telescope Compact Array is part of the Australia Telescope National Facility which is funded by the Commonwealth of Australia for operation as a National Facility managed by CSIRO. Observations reported in this paper were made under ATCA project codes C2914, C2965 and C3214.




\bibliographystyle{mnras}
\bibliography{1322cycle} 







\bsp	
\label{lastpage}
\end{document}